\documentclass[a4paper]{article}

\usepackage{amsmath,amsfonts,graphicx,epsfig,mathrsfs}
\newcommand{\R}{{\mathbb{R}}} 
\newcommand{\Z}{{\mathbb{Z}}} 
 
\newcommand{\C}{{\mathbb{C}}} 
\newcommand{\cx}{{{\mathbb{C}}^\times}}

\newcommand{\CP}{{\mathbb{C}}{{P}}}

\newcommand{\beq}{\begin{equation}} 
\newcommand{\eeq}{\end{equation}} 
\newcommand{\bea}{\begin{eqnarray}} 
\newcommand{\eea}{\end{eqnarray}} 
\newcommand{\ra}{\rightarrow}

\newcommand{\cd}{\partial} 
\newcommand{\wt}{\widetilde} 
 
\newcommand{\ol}{\overline} 
 
\newcommand{\ms}{{\sf M}} 
\newcommand{\mse}{{\sf M}^{\rm eq}} 
\newcommand{\mset}{\wt{\sf M}^{\rm eq}} 
\newcommand{\eps}{{\varepsilon}}

\newcommand{\nv}{{\bf n}} 
 
\newcommand{\sv}{{\bf S}}

\newcommand{\zv}{{\bf 0}} 
 
\newcommand{\DD}{\mathscr{D}}

\begin{document}

\title{Magnetic bubble refraction and quasibreathers in inhomogeneous
antiferromagnets}

\author{J.M. Speight\\
School of Mathematics, University of Leeds, Leeds LS2 9JT, UK
}

\date{}
\maketitle

\begin{abstract}
The dynamics of magnetic bubble solitons in a two-dimensional
isotropic antiferromagnetic spin lattice is studied, in the
case where the exchange integral $J(x,y)$ is position dependent.
In the near continuum regime, this system is described by the relativistic
$O(3)$ sigma model on a spacetime with a spatially inhomogeneous metric,
determined by $J$. 
The geodesic approximation is used to describe low energy soliton dynamics 
in this system: $n$-soliton motion is approximated by geodesic motion
in the moduli space $\ms_n$ of static $n$-solitons, equipped with the
$L^2$ metric $\gamma$. Explicit formulae for $\gamma$ for
various natural choices of $J(x,y)$ are obtained. From these it is shown
that single soliton trajectories experience refraction, with $J^{-1}$
analogous to the refractive index, and that this refraction effect
allows the construction of simple bubble lenses and bubble guides.
The case where $J$ has a disk inhomogeneity (taking the value $J_+$ outside
a disk, and $J_-<J_+$ inside) is considered in detail. It is argued
that, for sufficiently large $J_+/J_-$ this type of antiferromagnet
supports approximate quasibreathers: two or more coincident bubbles confined
within the disk which spin internally while their shape undergoes 
periodic oscillations with a generically incommensurate period.  
\end{abstract}

\section{Introduction}

The purpose of this paper is to analyze the dynamics of
bubble-like topological solitons arising in a two-dimensional
classical Heisenberg antiferromagnet where the exchange interaction
is isotropic, but not homogeneous. 
At each site $(i,j)\in\Z^2$ in a square lattice, one has a classical spin
vector
 $\sv_{ij}\in\R^3$, with 
$|\sv_{ij}|^2=s^2$, evolving according to the law
\beq\label{dag} 
\frac{d\sv_{ij}}{d\tau}=-\sv_{ij}\times\frac{\cd H}{\cd \sv_{ij}},
\qquad
H:=\sum_{i,j}J_{ij}\left[2s^2+\sv_{ij}\cdot(\sv_{i,j+1}+\sv_{i+1,j})\right]
\eeq
where $\tau$ is time and $J_{ij}$ are positive constants.
This is the usual planar isotropic Heisenberg antiferromagnet, but in
the case that the exchange integral $J$ has been made {\em inhomogeneous},
that is, position dependent. The experimental means of achieving this
$J$-inhomogeneity is left unspecified. One (perhaps naive) suggestion is
that a sample of one antiferromagnetic material could be enriched, in places,
with atoms from a different antiferromagnetic species. 
Note that since $J_{ij}>0$ for all $i,j$, $H$ is minimized when each spin 
anti-aligns with
its nearest neighbours, just as in the homogeneous case. 

It has been known for many years \cite{hal} that the dynamics of
the homogeneous spin lattice ($J_{ij}=J>0$, constant)
is described in the continuum limit by the $O(3)$ nonlinear sigma model.
In \cite{spebub} it was shown, by adapting the dimerization 
process of Komineas and Papanicolaou \cite{kompap}, that this result
generalizes in a natural way to the inhomogeneous problem considered
here. That is, the continuum limit of system (\ref{dag}) is described by
the nonlinear PDE
\beq
\label{oli}
\nv\times \DD\nv =\zv,\qquad
\DD\nv:=\frac{\cd^{2}\nv}{\cd t^2}-J(x,y)^2\left(\frac{\cd^2\nv}{\cd x^2}+
\frac{\cd^2\nv}{\cd y^2}\right)
\eeq
where $\nv:\R^{2+1}\ra\R^3$ has $|\nv|=1$, $t$ is a rescaled
time variable, and 
$J(x,y)$ is a continuum approximant of $J_{ij}$.
The relationship between $\nv$ and $\sv_{ij}$ is subtle. 
If we choose a lattice spacing $\eps>0$, assumed small, then
the spin dynamics associated with a solution $\nv(t,x,y)$ of (\ref{oli}) is
\beq\label{car}
\sv_{ij}(\tau)=(-1)^{i-j}s\nv(2\sqrt{2}s\eps\tau,i\eps,j\eps)+O(\eps).
\eeq
Roughly speaking, one should
imagine the field $\nv$ sampled on a square lattice of spacing
$\eps$, partitioned into
black and white sublattices, chess-board fashion. The white spins evolve
as $s\nv(t)$, the black as $-s\nv(t)$, where $t=2\sqrt{2}s\eps\tau$. 

Equation (\ref{oli}) is still the field equation of
the nonlinear $O(3)$ sigma model on $\R^{2+1}$, but now spacetime is
equipped with a spatially inhomogeneous Lorentzian metric,
\beq
\eta=dt^2-\frac{1}{J(x,y)^2}(dx^2+dy^2).
\eeq
Note that the static field equation is independent of $J$, so the usual 
Belavin-Polyakov lumps of the homogeneous system \cite{belpol}, 
which we shall call
bubbles in this context (in analogy with magnetic bubbles in ferromagnets), 
carry over unchanged to the inhomogeneous system. Given the relativistic
nature of (\ref{oli}), one can instantaneously boost these static bubbles
to obtain moving bubble solutions. However, due to the inhomogeneity
of $\eta$, such bubbles do not necessarily travel along straight lines.
A detailed study of the trajectory of a single bubble
interacting with various $J$-inhomogeneities was presented in \cite{spebub}.
 The main conclusion is that, under most circumstances, bubble
trajectories experience refraction, in analogy with geometric
optics, $J(x,y)^{-1}$ playing the role of the refractive index of the medium.
The incident and exit angles of a bubble crossing a domain wall 
$J$-inhomogeneity, for example, are related by Snell's law, and total
internal reflexion occurs if the impact is sufficiently oblique. 

In the current paper,
we shall briefly review this key 
refraction phenomenon, then go on to study some new dynamical problems
which were not considered in \cite{spebub}. 
Bubble refraction can be derived using only conservation laws
and the assumption that the scattering of a bubble from a
domain wall is approximately elastic \cite{spebub}. However, in this
paper our discussion of bubble dynamics will be entirely within
the gedoesic (or adiabatic)
approximation of
Manton \cite{man}, the standard approach to low-energy soliton dynamics
in field theories of this type. The geodesic approximation, 
 as applied to slow bubble dynamics, is set up
in section \ref{geoapp}.
This framework is then used in section \ref{domwal} to study
the interaction of
a single bubble with domain wall, trough and
disk 
$J$-inhomogeneities. Finally the rotationally equivariant
dynamics of $n$ coincident bubbles in a disk $J$-inhomogeneity is
studied in section \ref{cirlen}. Here a dynamical phenomenon entirely
absent from the homogeneous model, and with no optical analogue,
is found. A charge $n$ bubble set spinning about its symmetry axis can,
in some circumstances, undergo periodic shape oscillations. These
(approximate) solutions are spatially localized and, since the
period of the shape
oscillation need not be commensurate with the internal rotation period, 
in general, quasiperiodic in time. Hence they are {\em quasibreathers}.
By contrast, a charge $n$ bubble set spinning in the homogeneous model
simply spreads out indefinitely. The inhomogeneity of $J$ may thus act
to stabilize spinning bubbles.

\section{The geodesic approximation}
\label{geoapp}

Equation (\ref{oli})  is the variational equation for the action 
\beq
S=\frac{1}{2}\int dt\, dx\, dy\,
\sqrt{|\eta|}\, \cd_\mu\nv\cdot\cd_\nu\nv\, \eta^{\mu\nu}
=\int dt\, (T-V) 
\eeq
where
\beq
T=\frac{1}{2}\int dx\, dy 
\frac{1}{J(x,y)^2}\left|\frac{\cd\nv}{\cd t}\right|^2,\qquad
V=\frac{1}{2}\int dx\, dy
\left(\left|\frac{\cd\nv}{\cd x}\right|^2+
\left|\frac{\cd\nv}{\cd y}\right|^2\right)
\eeq
are identifed as the kinetic and potential energy functionals respectively.
Note that while $T$ depends on $J$, $V$ does not.
This is a consequence of the deep fact that the sigma model
potential energy (often called the harmonic map or Dirichlet energy)
on a space of dimension $2$
depends only on the conformal class of the spatial metric. In our
problem this metric is $g=J(x,y)^{-2}(dx^2+dy^2)$ which is conformal
to the Euclidean metric for any choice of $J(x,y)$.
It follows that Belavin's and
Polyakov's analysis of the {\em static} homogeneous
 model \cite{belpol} carries
over unchanged to the inhomogeneous system. There is a topological lower
energy bound
\beq
\label{belpolbou}
V[\nv]\geq 4\pi|n|,
\eeq
$n$ being the topological degree of the
map $\nv:\R^2\cup\{\infty\}\ra S^2$. It is useful to
 define complex coordinates 
$z=x+iy$ on the spatial plane and $u=(n_1+in_2)/(1-n_3)$ on the
target sphere (the latter being the image of $\nv$ under stereographic
projection from $(0,0,1)$ to the equatorial plane). Then the bound 
(\ref{belpolbou}) for $n\geq 0$ is attained if and only
if $u(z)$ is a rational map of algebraic degree $n$, that is
\beq
\label{nev}
u(z)=\frac{a_0+a_1z+\cdots+a_nz^n}{b_0+b_1z+\cdots+b_nz^n}
\eeq
where $a_i,b_i$ are complex constants,
 at least one of $a_n,b_n$ is nonzero, and the numerator and
denominator share no common roots. 
Since such maps minimize $V$ globally within their homotopy class,
they are automatically stable static solutions of the model.

Without loss of generality we may choose the
boundary value of $\nv$ at spatial infinity to
be $(0,0,1)$, that is $u(\infty)=\infty$, so that $b_n=0$
and hence $a_n\neq 0$. 
Since $a_n\neq 0$, we may divide the numerator and denominator
by $a_n$ and relabel $a_i/a_n\mapsto a_i$, $b_i/b_n\mapsto b_i$, so 
that
the general static solution is
\beq\label{min}
u(z)=
\frac{a_0+a_1z+\cdots+a_{n-1}z^{n-1}+z^n}{b_0+b_1z+\cdots+b_{n-1}z^{n-1}}.
\eeq
This is uniquely specified by the $2n$ complex constants
$a_0,\ldots,b_{n-1}$. Hence
the moduli space (or parameter space) of degree $n$ static
solutions, $\ms_n$, may be identifed with an open subset of 
$\C^{2n}$, namely the complement of the complex codimension $1$ algebraic
variety on which the numerator and denominator have common roots.
For example, 
the general degree 1 bubble is
\beq\label{u0}
u(z)=\chi^{-1}e^{-i\psi}(z-w),
\eeq
$\chi\in(0,\infty)$, $\psi\in[0,2\pi]$ and $w=w_1+iw_2\in\C$ being constants 
interpreted as the bubble's width, internal phase and position respectively.
Hence $\ms_1$ is diffeomorphic to $\cx\times\C$, where
$\cx=\C\backslash\{0\}$. In general $\ms_n$ has
real dimension $4n$, and may be thought of as the configuration space of
$n$ unit bubbles. However, it is not diffeomorphic to $(\ms_1)^n$, and
its parameters cannot globally be identified with the positions, widths and
orientations of $n$ separate bubbles. 

Following Ward \cite{war} and Leese \cite{lee}, we may study the low energy
dynamics of $n$ bubbles within the geodesic approximation of Manton
\cite{man}. The idea is that $\ms_n$, the $4n$ dimensional space 
 of static degree $n$ solutions, is the flat valley 
bottom in the space of all degree $n$ maps $\R^2\ra S^2$, 
on which $V$ attains its
minimum value of $4\pi n$. 
Consider the Cauchy problem in which the constituent bubbles of
a static $n$-bubble are set off in relative motion with small speeds.
Then the initial field lies in $\ms_n$, the initial velocity field is 
tangential to $\ms_n$, and the system has only
 a small amount
of kinetic energy. Hence its subsequent motion in field configuration
space is confined close to $\ms_n$ by
conservation of energy $E=T+V$ (note that $T$ is strictly positive).
This suggests that a collective coordinate
approximation wherein the motion is constrained to $\ms_n$ for all time is 
sensible. This is often called an adiabatic approximation, since we
assume that at each time the field configuration is well approximated by
 a static solution, but that the parameters in this solution may vary slowly
with time.

Let $q^1,\ldots,q^{4n}$ be an arbitrary real coordinate system on $\ms_n$
(for example, the real and imaginary parts of 
$a_0,\ldots,a_{n-1},b_0,\ldots,b_{n-1}$), and let us denote by
$u_0(z;q)$ the static $n$-bubble corresponding to $q=(q^1,\ldots,q^{4n})$.
Then the adiabatic approximation asserts that 
\beq\label{colco}
u(z,t)=u_0(z;q(t)).
\eeq
We now substitute (\ref{colco}) into $S$, and 
obtain variational equations for
$q$. Note that $V\equiv 4\pi n$, constant, for all fields of the
form (\ref{colco}), and $T$ is quadratic in time derivatives, 
so the action reduces to (Einstein summation convention applied)
\beq
S=\frac{1}{2}\int dt\, (\gamma_{ij}\dot{q}^i\dot{q}^j-4\pi n),\qquad
\gamma_{ij}(q):=\int\frac{dx dy}{J(x,y)^2}
\frac{4}{(1+|u_0(z;q)|^2)^2}\frac{\cd u_0}{\cd q^i}
\frac{\cd\bar{u}_0}{\cd q^j}.
\eeq
This is the action for geodesic motion on $\ms_n$ with respect to the
metric $\gamma=\gamma_{ij}(q)dq^idq^j$. Hence we
expect $u(z,t)$
to be well
 approximated by geodesic motion in $(\ms_n,\gamma)$, at least for low
speeds. It should be emphasized that this does {\em not} mean that
individual bubbles follow geodesics in physical space
$(\R^2,g)$, even in the case $n=1$. Indeed, the motion of $2$-bubbles
even in Euclidean space (i.e.\ even in the homogeneous model) is
highly nontrivial \cite{war,lee}. For example, two identical bubbles
fired directly at one another do not pass through one another unchanged,
or  achieve some minimum separation then recede along their line of approach,
as one might expect. Rather, they coalesce to form a ringlike coincident
$2$-bubble, then break apart and recede along a line 
{\em perpendicular} to their
line of approach. This behaviour is generic to planar topological
solitons in relativistic field theories, and is well-understood within
the geodesic approximation \cite{rub}. The conceptual framework and 
validity of the geodesic 
approximation are discussed at length in \cite{mansut}. 

The metric $\gamma$, and hence the dynamics, depends
strongly on $J(x,y)$. The main task in this approach
to topological
soliton dynamics is to calculate this metric. Even in the
degree $1$ case this is impossible to compute explicitly unless 
$J$ takes a particularly simple, symmetric form. 
Nonetheless, much 
qualitative information about $\gamma$ can be deduced, and this can lead
to a good understanding of the dynamics of a single bubble even in the
absence of explicit formulae. In the case of higher degree, $n\geq 2$,
the moduli space is too large for the general geodesic problem to be 
tractable, even if $\gamma$ is known explicitly. To make progress
one must reduce
the dimension by identifying totally geodesic submanifolds. This amounts
to imposing symmetry constraints on the initial data.

The metric in the case $n=1$ has a particularly simple form.
If $J$ is constant, we are studying the standard
sigma model, and it is found that $\chi$ and $\psi$ are frozen by infinite
inertia ($\gamma_{\chi\chi}=\gamma_{\psi\psi}=\infty$) 
and $\gamma=4\pi J^{-2} dw\, d\bar{w}$, so bubbles
just travel at constant velocity \cite{war}.  At the other extreme,
if $J(x,y)=1+x^2+y^2$, we effectively have the sigma model on a round
two-sphere, and the bubble dynamics is much richer \cite{spe}. This choice
has $J$ unbounded, which is presumably unphysical, however. Assuming
that $J$ remains bounded,  $\chi$ and $\psi$ are frozen by
an essentially identical argument to Ward's. Without loss of 
generality, we can set $\psi=0$. The width $\chi$ remains a free, but frozen,
parameter. The metric on $\ms_1^\chi$,
 the width $\chi$, phase 0 leaf of $\ms_1$, is then
\beq
\gamma=f_\chi(w)(dw_1^2+dw_2^2),
\eeq
where
\beq\label{fgeom}
f_\chi(w)=
\int\frac{dz\, d\bar{z}}{J(z,\bar{z})^2}\, 
\frac{4\chi^{-2}}{(1+\chi^{-2}|z-w|^2)^2}
=\int \frac{4\,  du\, d\bar{u}}{(1+|u|^2)^2}\, \frac{1}{J(\chi u+w)^2},
\eeq
the integral of $J(\chi u +w)^{-2}$ over the standard unit sphere on
which $u$ is
a stereographic coordinate.  Note that
\beq
\lim_{\chi\ra 0} \gamma=\frac{4\pi}{J(w)^2}(dw_1^2+dw_2^2)=4\pi g.
\eeq
In the limit of vanishing bubble width, therefore, the
trajectories of isolated bubbles
tend to geodesics, not just in $\ms_1^\chi$, but also in the
physical plane. 
One may regard the metric $\gamma$ on $\ms_1^\chi$ as a smeared out version
of $g$, due to the
bubble's finite core size. 

\section{Bubble refraction}
\label{domwal}

The simplest $J$-inhomogeneity of all is
the domain wall. Let us assume that $J$ depends only on $x$, is constant
outside a small neighbourhood $(-\delta,\delta)$ of 
$x=0$ and rises monotonically from
$J_-$ to $J_+$ as $x$ traverses this neighbourhood. Such a $J(x)$ might arise
from enriching one end of an antiferromagnet but not the other, for example.
It follows immediately from (\ref{fgeom}) that the conformal
factor $f_\chi(w)$ of the metric $\gamma$ on $\ms_1^\chi$ depends only
on $w_1$, is monotonically {\em decreasing} and has
\beq\label{erh}
\lim_{w_1\ra\pm\infty}f_\chi(w_1)=\frac{4\pi}{J_\pm^2}.
\eeq
Our analysis of the interaction of the bubble with the domain wall
will use only these properties of $f_\chi$. An explicit formula
for $f_\chi$ can be obtained in the limit of vanishing domain wall
width $\delta\ra 0$.
If we idealize the domain wall by a step function,
\beq
J(x)=\left\{\begin{array}{ll} J_+ & x\geq 0\\
J_- & x<0.\end{array}\right. ,
\eeq
we find from (\ref{fgeom}),
 and some elementary spherical geometry
\cite{spebub}, that
\beq\label{f}
f_\chi(a)=\frac{2\pi}{J_-^2}\left[1-\frac{a_1}{\sqrt{\chi^2+a_1^2}}\right]
+\frac{2\pi}{J_+^2}\left[1+\frac{a_1}{\sqrt{\chi^2+a_1^2}}\right].
\eeq
Note that $f_\chi$ decreases smoothly from $4\pi/J_-^2$ to $4\pi/J_+^2$,
as expected.

\begin{figure}[htb]
\centering
\includegraphics[scale=0.35]{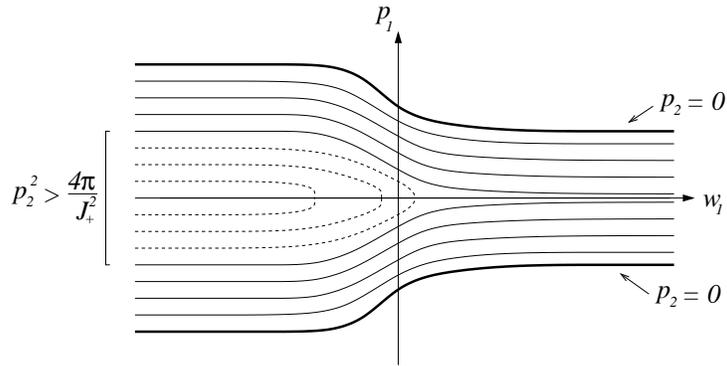}
\caption{\sf 
Projected phase portrait of the geodesic flow in $\ms_1^\chi$
for a domain wall $J$-inhomogeneity. The exchange integral on the right
of the wall, $J_+$, exceeds that on the left, $J_-$. Trajectories are bounded
between the upper and lower bold curves, for which $p_2=0$. 
Dashed trajectories, which have 
$\frac{4\pi}{J_+^2}<p_2^2<\frac{4\pi}{J_-^2}$, correspond to bubbles 
undergoing total internal reflexion.
}
\label{fig1}
\end{figure}

To analyze geodesic flow in $\ms_1^\chi$,
it is helpful to use the Hamiltonian formalism. The position coordinates
$w_1,w_2$ have canonically conjugate momenta 
\beq
\label{oev}
p_1=f(w_1)\dot{w}_1,\qquad
p_2=f(w_1)\dot{w}_2,
\eeq
and the flow is generated by the Hamiltonian
\beq
\label{enn}
H=\frac{1}{2f(w_1)}(p_1^2+p_2^2).
\eeq
Both $H$ and $p_2$ are integrals of the motion. Since geodesic trajectories 
are independent of initial speed, we may assume, without loss of generality,
that $H=\frac{1}{2}$. Trajectories now fall into a one-parameter family
labelled by the conserved momentum $p_2$, which given (\ref{enn}) must
satisfy $p_2^2< \sup_{w_1\in\R}f(w_1)=4\pi/J_-^2$. The image of a
momentum $p_2$ 
trajectory under projection to the $(w_1,p_1)$ plane in phase space lies
in the level set $f(w_1)-p_1^2=p_2^2$. 
A sketch of these (projected) trajectories is presented in figure
\ref{fig1}.
They are confined to  
the bottle shaped region bounded by the curves $p_1=\pm\sqrt{f(w_1)}$
whose union is the level set with $p_2^2=0$. These curves themselves
correspond to a bubble hitting the domain wall with incidence angle 0
(i.e.\ orthogonal to the wall) and travelling straight through it,
either left to right (upper curve) or right to left (lower). Two types
of trajectory are evident. If $0\leq p_2^2<4\pi/J_+^2$, then $\dot{w}_1$
never vanishes and the bubble passes through the wall. We 
call these refracted trajectories. If $4\pi/J_+^2<p_2^2<4\pi/J_-^2$, the
trajectories have $\lim_{t\ra\pm\infty}w_1(t)=-\infty$, so the bubble
approaches the wall from the left and is reflected by it. We call
these reflected trajectories. Let $\theta(t)$ be the angle between the
bubble's velocity and the $x$ axis, so $\dot{w}=|\dot{w}|e^{i\theta(t)}$.
Then
\beq\label{eed}
\sin\theta(t)=\frac{\dot{w}_2}{\sqrt{\dot{w}_1^2+\dot{w}_2^2}}=
\frac{p_2}{\sqrt{p_1^2+p_2^2}}=\frac{p_2}{\sqrt{f(w_1(t))}}.
\eeq
Hence for a refracted trajectory incident on the domain wall from the
left,
\beq\label{S}
\frac{\sin\theta(\infty)}{\sin\theta(-\infty)}=\frac{f(-\infty)}{f(\infty)}
=\frac{J_+^{-1}}{J_-^{-1}}
\eeq
so the incident and exit angles of the trajectory, $\theta(-\infty)$ and 
$\theta(\infty)$, are related by
Snell's law of refraction, with $J^{-1}$ identified with the refractive
index. 
If $\sin\theta(-\infty)>J_-/J_+$, then $p_2^2>4\pi/J_+^2$ and
total internal reflexion occurs.

It is interesting to make a similar qualitative analysis in the
case of a straight trough $J$ inhomogeneity. Assume that
$J$ is again independent of $y$, that $J(x)=J_1$ for $|x|\geq 1+\delta$,
$\delta>0$, small; that
$J(x)=J_2<J_1$ for $|x|\leq 1-\delta$; and that
$J$ interpolates monotonically
between the constant values $J_1,J_2$ in the narrow transition intervals
$(-1-\delta,-1+\delta)$ and $(1-\delta,1+\delta)$. Given Snell's law for
domain walls (\ref{S}), one expects this to be a bubble guide: 
we have an infinite strip in which the refractive
index is $1/J_2$, greater than the refractive index $1/J_1$
in the surrounding region, so a bubble initially in and moving roughly along
the strip should be trapped in the strip by total internal reflexion,
just as a light ray is confined within a fibre-optic cable. 

The key point is, once more, to understand the conformal factor $f_\chi(w)$
of the metric on $\ms_1^\chi$. Clearly $f_\chi$ depends only on $w_1$ and
tends to $4\pi/J_1^2$ as $w_1\ra\pm\infty$. In general, it
is hump shaped, with a single critical point, a global maximum.
If we assume $J(w_1)$ is
even, then so is $f_\chi(w_1)$, and this maximum, which is
always less than $4\pi/J_2^2$, occurs at $w_1=0$. Again, in the sharp
wall limit, $\delta\ra 0$, we can find an explicit expression for
$f_\chi(w_1)$. The point is that $J(w_1)$ is piecewise constant in this 
limit, so from (\ref{fgeom}), we need only compute the area of the
region on $S^2$ whose stereographic image is the vertical strip
$\{u\in\C\: :\: -1<{\rm Re}\, (\chi u +w)<1\}$. This is easily
deduced from the sharp domain wall calculation in \cite{spebub}
(which led to formula (\ref{f})), yielding
\beq
\label{sag}
f_\chi(w_1)=\frac{4\pi}{J_1^2}+2\pi
\left(\frac{1}{J_2^2}-\frac{1}{J_1^2}\right)
\left[\frac{w_1+1}{\sqrt{(w_1+1)^2+\chi^2}}-\frac{w_1-1}{\sqrt{(w_1-1)^2+
\chi^2}}\right].
\eeq
Note that this function has the qualitative features predicted.

\begin{figure}[htb]
\centering
\includegraphics[scale=0.35]{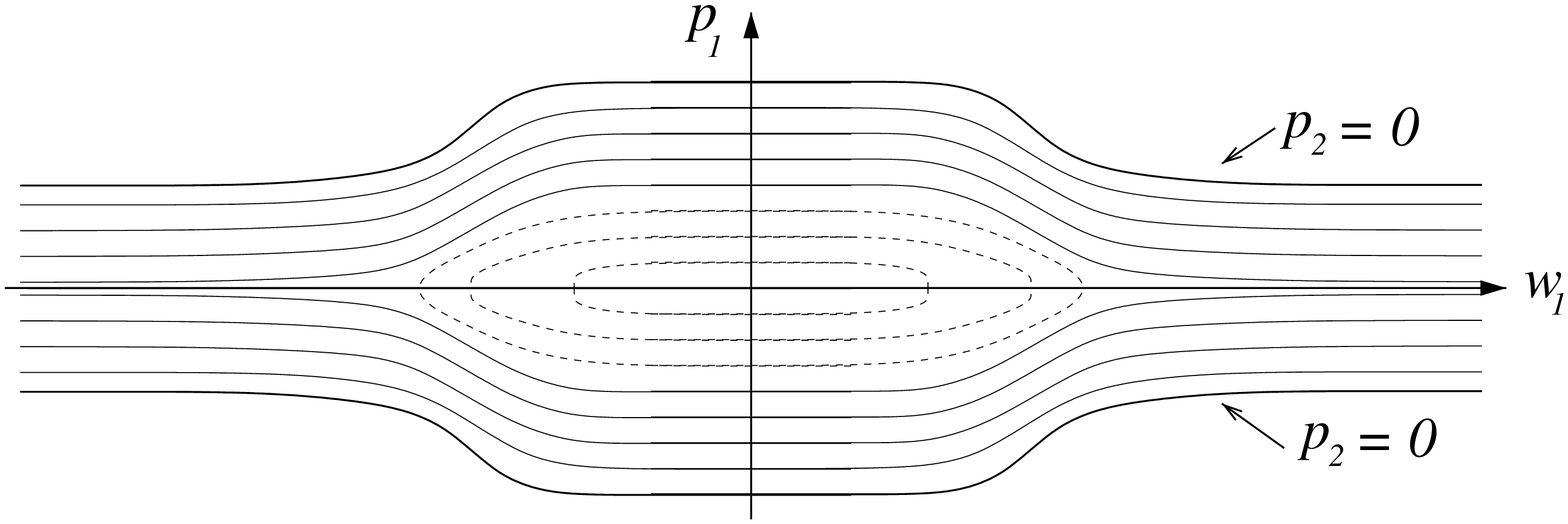}
\caption{\sf 
Projected phase portrait of the geodesic flow in $\ms_1^\chi$
for a trough $J$-inhomogeneity: the exchange integral is
$J_1$ outside a vertical strip, within which it falls to $J_2<J_1$.
 The closed dashed trajectories, which have 
$p_2^2>\frac{4\pi}{J_1^2}$, correspond to bubbles being
guided along the trip by total internal reflexion.
}
\label{fig:ood}
\end{figure}

The Hamiltonian discussion of geodesic flow in $(\ms_1^\chi,\gamma)$
is similar to the domain wall case. Once again we choose
$H=\frac{1}{2}$ and sketch the level curves 
$
f_\chi(w_1)-p_1^2=p_2^2
$
in the $(w_1,p_1)$ plane for $0\leq p_2^2<f_{\rm max}=f_\chi(0)$.
Again, two regimes emerge separated by the case $p_2^2=p_*^2=4\pi/J_1^2$,
see figure \ref{fig:ood}.
For $p_2<p_*$ the trajectories are unbounded and $\dot{w}_1$ never vanishes.
These correspond to a bubble hitting the trough from outside and passing 
right through
it, being refracted twice in the process, so that its initial and final
velocities coincide. For $p_2>p_*$ the (projected)
trajectories are bounded and periodic. These correspond to a bubble being 
directed along the direction of the trough by total internal reflexion.
The actual bubble trajectory in the physical plane is not periodic, of 
course, since it drifts uniformly 
in the $y$-direction ($\dot{w}_2$
never vanishes; it is {\em not} constant, however). Note that even
in this trapped regime, the bubble trajectory is not necessarily
confined within the trough itself. Indeed, given any $R>0$,
there exists a trapped geodesic for which  $\max_t|w_1(t)|>R$. In this
sense, the behaviour of bubble trajectories is quite different from
that of light rays in geometric optics.

The bubble guide effect does not depend on the trough being straight.
The case of a circular trough (where the value of $J$ is
suppressed in an annulus) was considered in \cite{spebub}, the results being
qualitatively similar.

\begin{figure}[htb]
\centering
\includegraphics[scale=0.5]{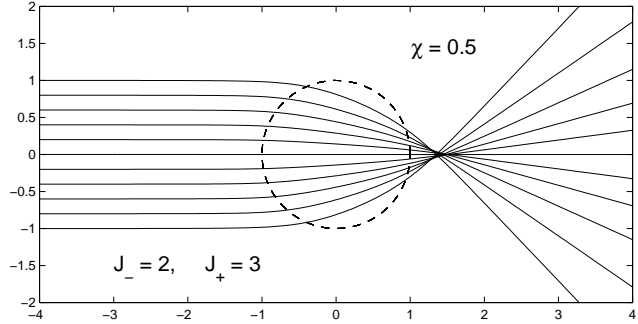}
\caption{\sf 
Focussing of parallel bubble trajectories by a disk $J$-inhomogeneity. The
exchange integral is lower ($J_-=2$) within the disk than in the surrounding
plane ($J_+=3$).}
\label{fig:ing}
\end{figure}

Another interesting effect reminiscent of geometric optics is
the phenomenon of bubble trajectory focussing by a circular ``lens''.
Imagine $J$ depends only on $|z|$, so that $J(|z|)=J_-$ for 
$|z|\leq 1-\delta$, $J(|z|)=J_+>J_-$ for $|z|>1+\delta$, interpolating
smoothly and monotonically between these values in the narrow annulus
$1-\delta<|z|<1+\delta$. We have a disk where the refractive index
$1/J_-$ exceeds that of the surrounding medium, so we expect single
bubble trajectories to be refracted by this inhomogeneity like light
rays passing through a circular lens. 
It follows immediately from (\ref{fgeom}) that the conformal factor in the
metric depends only on $|w|$.
In the sharp wall limit $\delta\ra 0$, it can be explicitly computed,
\beq\label{dbe}
f_\chi(|w|)=\frac{2\pi}{J_-^2}\left[1-\frac{\chi^2+|w|^2-1}{\sqrt{\chi^4+
2\chi^2(|w|^2+1)+(|w|^2-1)^2}}\right]+
\frac{2\pi}{J_+^2}\left[1+\frac{\chi^2+|w|^2-1}{\sqrt{\chi^4+
2\chi^2(|w|^2+1)+(|w|^2-1)^2}}\right].
\eeq
The geodesic problem in $\ms_1^\chi$ may easily be solved numerically,
see figure \ref{fig:ing}. Note that parallel incident trajectories are
approximately focussed by the lens, as expected.

\section{Quasibreathers}
\label{cirlen}

All the $J$-inhomogeneities we have considered enjoy some kind
of symmetry, either translational (domain walls and troughs) or rotational
(disks).
In this section, we will continue to impose rotational invariance, so 
$J=J(|z|)$. 
However, 
our emphasis will be very different: we will study rotationally
equivariant $n$-bubble dynamics for $n\geq 2$. This corresponds to
$n$ coincident magnetic bubbles sitting, for all time, at the
symmetry origin $z=0$. This is dynamically interesting because
for $n\geq 2$ the $n$-bubble width is no longer a frozen parameter.
The $n$-bubble can spin internally and spread out or
sharpen. So the dynamics of interest is no longer bubble transport, but 
rather the {\em internal} 
motion of the $n$-bubble. We will again
model this within the geodesic approximation.

First note that, if $J=J(|z|)$, there is a natural isometric action of 
$U(1)$ on $(\ms_n,\gamma)$, given by 
\beq\label{ckc}
u(z)\mapsto e^{-ni\alpha}u(e^{i\alpha}z). 
\eeq
The fixed point set
of this action is necessarily a totally geodesic submanifold of
$(\ms_n,\gamma)$, which we shall call $(\mse_n,\gamma|)$, where
$\gamma|$ is the induced metric. In fact \cite{macspe},
$\mse_n$ is, for all $n$, diffeomorphic to $\cx$, since it consists
of the rational maps
\beq\label{thi}
u(z)=cz^n,\qquad c\in\cx.
\eeq
Since $U(1)$ equivariance is a valid symmetry reduction of the
full field equation (\ref{oli}), one may regard geodesic flow in
$(\mse_n,\gamma|)$ either as a low energy approximation of the
symmetry-reduced field equation, or as a symmetry reduction of the
general geodesic approximation. 

The geometry of $\mse_n$ is most easily understood by studying the
lift $\wt{\gamma}$ of $\gamma|$ to the $n$-fold cover $\mset_n$
of $\mse_n$ \cite{macspe}. This is again diffeomorphic to $\cx$, and the
covering projection is
\beq\label{ock}
\pi:\mset_n\ra\mse_n,\qquad
\pi(\rho)=\rho^{-n}.
\eeq
Hence, the $n$-bubble corresponding to $\rho\in\cx$ is
$u(z)=(z/\rho)^n$. Clearly any of the $n$ points $\rho e^{2\pi ik/n}$
in $\mset_n$, $k=0,1,\ldots,n-1$,
give the same $n$-bubble. We may think of $|\rho|$
as the $n$-bubble width, and the argument of $\rho$, modulo
$2\pi/n$, as being an internal phase. The potential energy density
of an $n$-bubble is rotationally symmetric, vanishes
at $z=0$ and $|z|\ra\infty$,  and attains its maximum 
at 
\beq\label{uph}
|z|=|\rho|\left(\frac{n-1}{n+1}\right)^\frac{1}{2n}.
\eeq
The lifted metric on $\mset_n$ is
\beq\label{erc}
\wt{\gamma}=f(|\rho|)d\rho d\ol{\rho},\qquad
f(|\rho|)={8\pi n^2}\int_0^\infty\, dr\,
\frac{r^{2n+1}}{(1+r^{2n})^2J(|\rho|r)^2}.
\eeq
In the homogeneous case, $J(|z|)=J_0$, constant, the metric is
$\wt{\gamma}=C_nd\rho d\ol\rho$
 where $C_n$ is a constant
depending on $n$, so that $(\mset_n,\wt{\gamma})$ is actually
isometric to the {\em Euclidean} punctured plane, and the
lifted geodesics are simply straight lines traversed at constant speed.
 It follows that all
geodesics have $|\rho|\ra\infty$ either as $t\ra\infty$ or as $t\ra-\infty$,
and, except in the case where $\rho(t)$ is a half ray hitting the
missing point $\rho=0$, both. The half rays correspond (when oriented
inwards) to an $n$-bubble with no angular momentum collapsing to form
a singularity in finite time. The straight lines missing $\rho=0$
correspond to $n$-bubbles with angular momentum which spread out
indefinitely as $|t|\ra\infty$. In total, each spin $\nv(z)$ in
such a bubble executes
$n/2$ complete rotations about the $n_3$ axis.

The behaviour is rather more interesting if we allow $J(|z|)$ to be 
non-constant. The ``spherical'' case $J(|z|)=1+|z|^2$ was studied in detail
in \cite{macspe}. One nice, and rather unexpected, fact in this case is
that the volume of $\mse_n$, which turns out to be finite, is actually
independent of $n$. One can isometrically embed $\mse_n$ in $\R^3$ as
a surface of revolution, and this surface has conical singularities of 
deficit angle $2\pi(1-n^{-1})$ at $c=0$ and $c=\infty$. However, this
case has $J$ unbounded, which is not physically reasonable. We
shall here consider the circular lens case,
 where $J$ is suppressed in a disk centred
on the origin. For simplicity, we will use the sharp wall limit,
that is
\beq\label{ute}
J(|z|)=\left\{\begin{array}{ll}
J_+& |z|>1\\
J_-& |z|\leq 1\end{array}\right.
\eeq
with $J_+>J_-$. 
We saw in the last section that the metric on $\ms_1^\chi$ could be
computed explicitly for this $J(|z|)$. We emphasize that this is {\em not}
the metric of interest in the present context, however.
Recall that position in $\ms_1^\chi$ is specified by the bubble's 
position in the physical plane, and the bubble width $\chi$ is a frozen
parameter. By contrast, position in $\mse_n$ is specified by the
$n$-bubble's width and internal phase, and its location in the physical
plane is fixed.
In \cite{spebub} an interesting bifurcation in the geodesic flow 
on $\ms_1^\chi$ was observed
when $\chi$ is varied: as $\chi$ increases through a critical
value, a pair of periodic geodesics appears in which the lump
orbits in a circle centred on $z=0$, one just inside the lens, one just
outside.  We will see that there is a geometrically
similar bifurcation in the geodesic flow in $\mset_n$,
but its physical meaning is now completely different, and the bifurcation
parameter is $J_+/J_-$, not the bubble width (now a dynamical variable).

\begin{figure}[htb]
\centering
\includegraphics[scale=0.6]{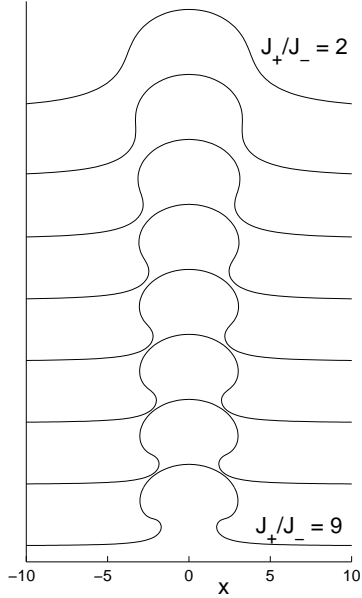}
\caption{\sf Generating curves for $\mset_2$ isometrically embedded as
a surface of revolution in $\R^3$ in the case of a sharp disk
$J$-inhomogeneity with $J_-/J_+=2,3,\ldots,9$ (top to bottom). The
surfaces are swept out as these curves are rotated about the vertical axis
$x=0$. Note that a neck develops as $J_+/J_-$ becomes large.
}
\label{fig:tle}
\end{figure}

Assuming $J$ is given by (\ref{ute}), one sees that
\beq\label{lit}
f(|\rho|)=\frac{2\pi n^2}{J_-^2}\left\{
\frac{\pi}{n^2\sin \pi/n}-\left(1-\frac{J_-^2}{J_+^2}\right)
\int_0^{|\rho|^2}ds\frac{s^{n-2}}{(1+s^n)^2}\right\}.
\eeq
Since the integrand is rational, one can, in principle, write down
$f$ explicitly for any choice of $n$. The expressions are not
terribly enlightening however, even for small $n$, so we will omit them.
More useful is the fact that, provided $J_+/J_-$ is not too large, 
$\mset_n$ can be isometrically embedded as a surface of revolution
in $\R^3$. 
Generating curves for these surfaces in the case $n=2$, 
generated by the method described in
\cite{macspe}, are depicted in figure \ref{fig:tle}.
If $J_+/J_-=1$ the surface is just a (punctured) plane,
swept out by rotating a straight half-line. As $J_+/J_-$ increases,
the generating curve becomes hump-shaped, until eventually, it forms a
bulge above
 a narrower neck. At this point, a bifurcation in the geodesic flow
occurs. A pair of periodic geodesics with $|\rho|$ constant appears, and
there are quasiperiodic (and periodic) geodesics which remain confined
within the bulge. These correspond to quasibreathers. As $J_+/J_-$
increases further, the isometric embedding is lost, but the 
existence of quasibreather geodesics persists. 

\begin{figure}[htb]
\centering
\begin{tabular}{cc}
(a)&(b)\\
\includegraphics[scale=0.4]{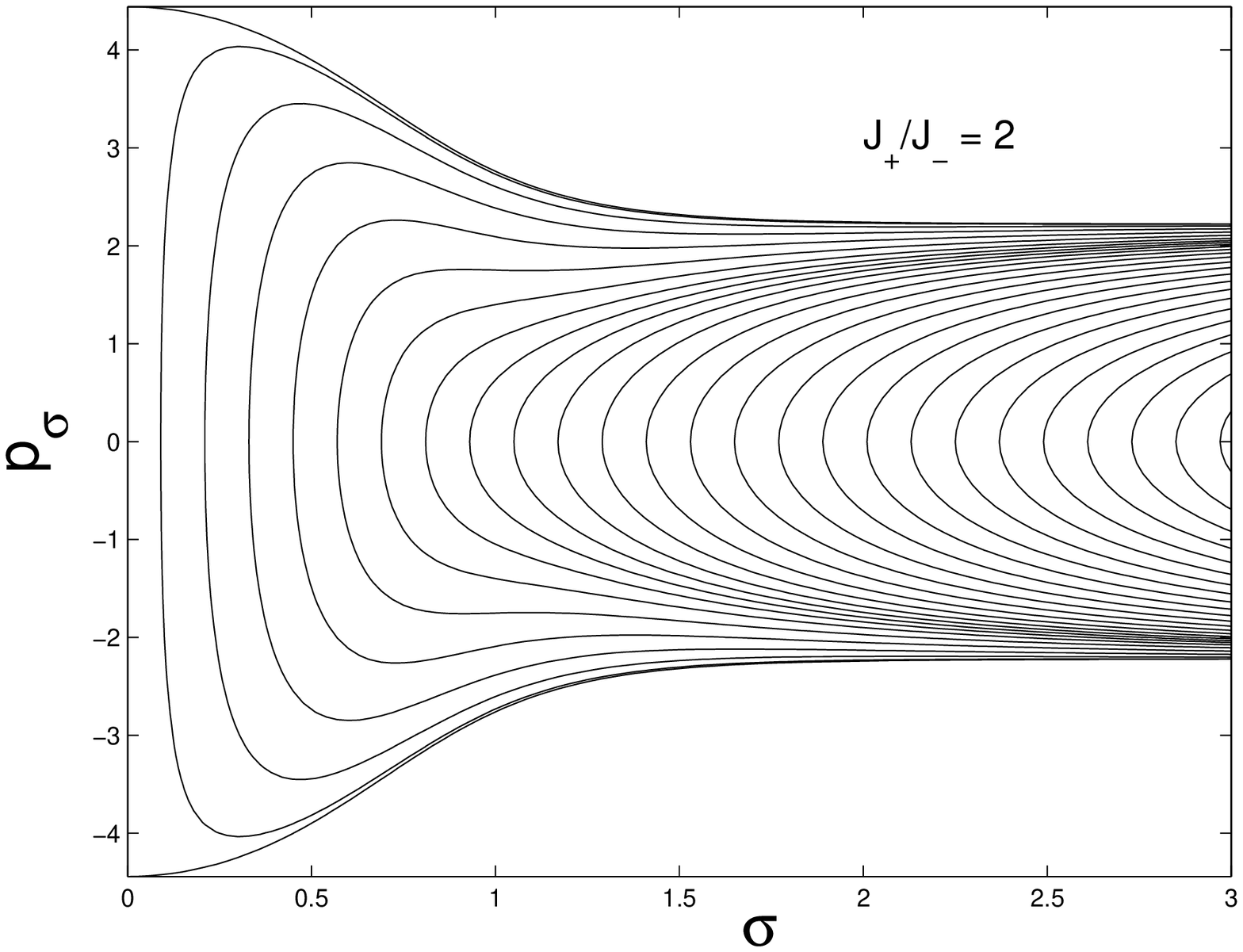}&
\includegraphics[scale=0.4]{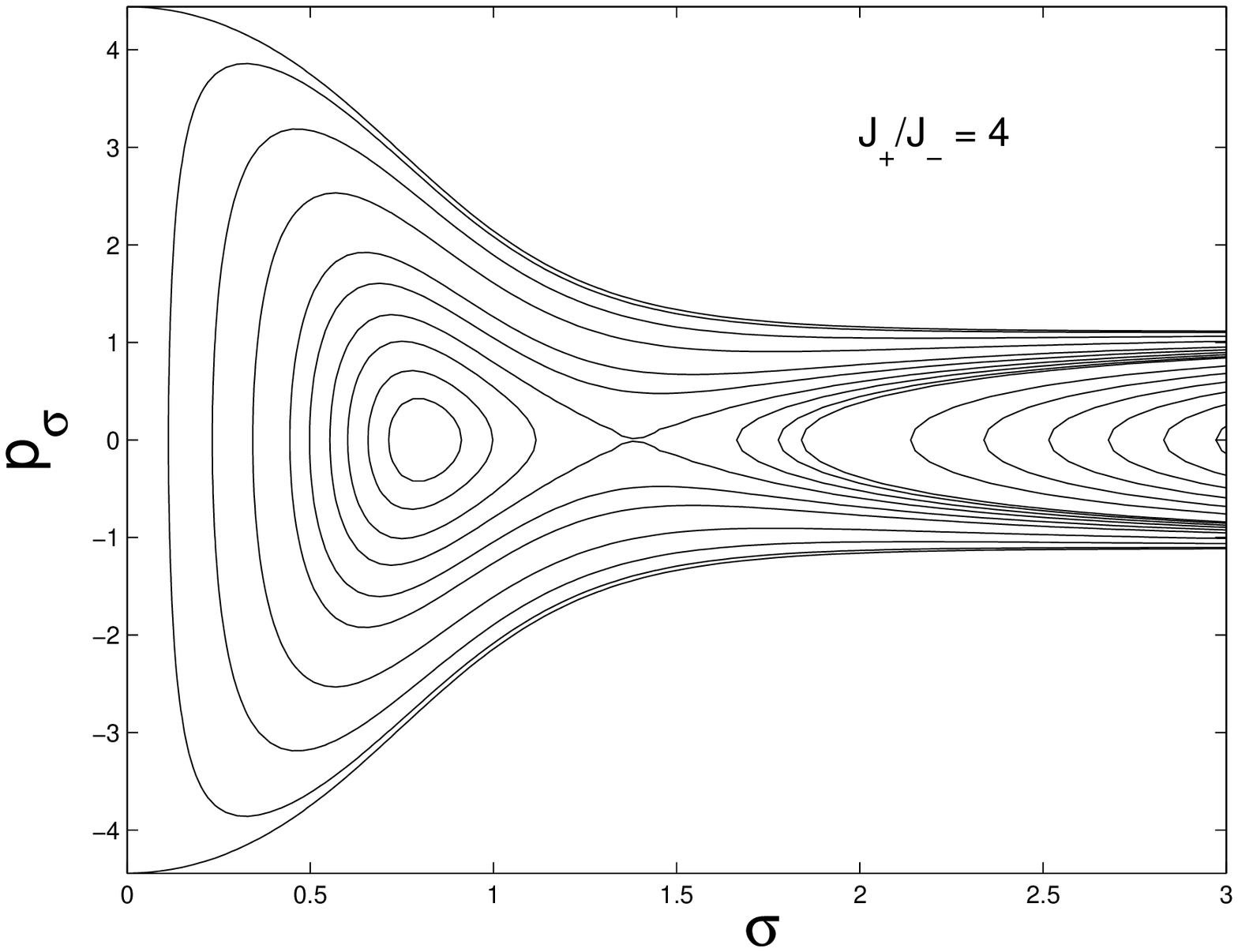}
\end{tabular}
\caption{\sf Phase portraits of the geodesic flow in $\mset_2$ for
(a) $J_+/J-=2$ and (b) $J_+/J_-$. Note the island of periodic orbits
for the higher value of $J_+/J-$.
}
\label{fig:diw}
\end{figure}

To confirm this visual reasoning, let us again consider the phase portrait
of the geodesic flow. Introducing polar coordinates $(\sigma,\psi)$
on $\mset_n$ by $\sigma e^{i\psi}:=
\rho$, the flow is generated by Hamiltonian
\beq
H=\frac{1}{2f(\sigma)}\left(p_\sigma^2+\frac{p_\psi^2}{\sigma^2}\right),
\eeq
the conjugate momenta being
\beq\label{ars}
p_\sigma=f(\sigma)\dot\sigma,\qquad
p_\psi=\sigma^2 f(\sigma)\dot\psi.
\eeq
Both $H$ and $p_\psi$ are integrals of the motion, and we may choose
$H=\frac{1}{2}$ without loss of generality. The trajectories, after
projection to the $(\sigma,p_\sigma)$ half-plane (note $\sigma>0$)
are given by the level sets
\beq\label{ean}
\sigma^2(f(\sigma)-p_\sigma^2)=p_\psi^2
\eeq
for $p_\psi^2\in [0,\infty)$. These level sets are plotted for
$J_+/J_-=2$ and $J_+/J_-=4$ in the case $n=2$ in figure \ref{fig:diw}.
Note that as $J_+/J_-$ exceeds a critical value (around $2.89$ for $n=2$), 
an island of periodic orbits appears in the phase portrait. 
This is due to the formation of a local maximum in $\sigma^2f(\sigma)$ and
hence, given the asymptotic behaviour $\sigma^2 f(\sigma)\sim 4\pi J_+^{-2}
\sigma^2$ as $\sigma\ra\infty$, due to the formation of a {\em pair}
of critical points of $\sigma^2 f(\sigma)$. These points have a nice
geometric interpretation, which we will now describe. First note that
\beq\label{oul}
\frac{d\:}{d\sigma}\sigma^2f(\sigma)=0\quad\Leftrightarrow\quad
\Xi(\sigma):=1+\frac{\sigma f'(\sigma)}{2f(\sigma)}=0.
\eeq 
Now, the necessary and sufficient condition for $\mset_n$ to have
an isometric immersion as a surface of revolution in $\R^3$ is 
that $-1\leq\Xi(\sigma)\leq 1$ for all $\sigma$, see \cite{macspe}.  
In this case, if we imagine the surface as being generated by rotating
a generating curve in the $x_1x_2$ plane about the $x_3$ axis (as depicted 
in figure \ref{fig:tle}), then $\Xi(\sigma)=\cos\xi(\sigma)$, where
$\xi$ is the angle between the tangent to the generating curve and the $x_1$
axis. Hence critical points of $\sigma^2f(\sigma)$ are associated with
points where the tangent to the generating curve is vertical, and a pair
of such points develops precisely when the surface develops a pinched
neck. This confirms our visual intuition that the island of closed
trajectories appears when the pinched neck forms in $\mset_n$.

\begin{figure}[htb]
\centering
\includegraphics[scale=0.5]{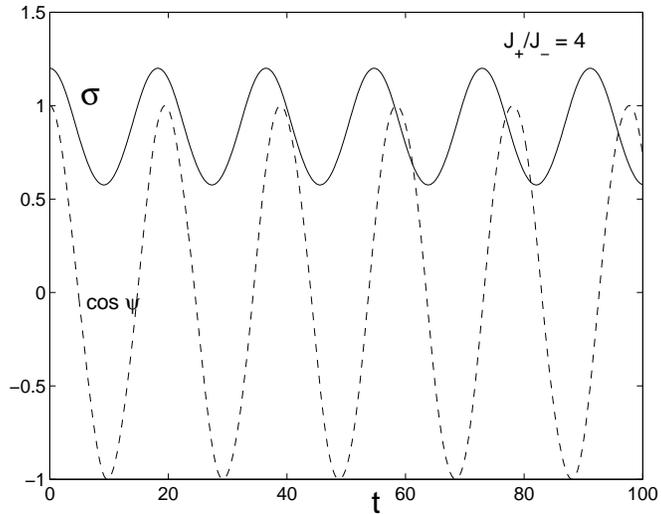}
\caption{\sf The time dependence of the width $\sigma$ (solid curve)
and internal
phase $\psi$ (dashed curve)
of a typical degree $2$ quasibreather, in a disk
inhomogeneity with $J_+/J_-=4$.}
\label{fig:dnt}
\end{figure}

\begin{figure}[htb]
\centering
\includegraphics[scale=0.5]{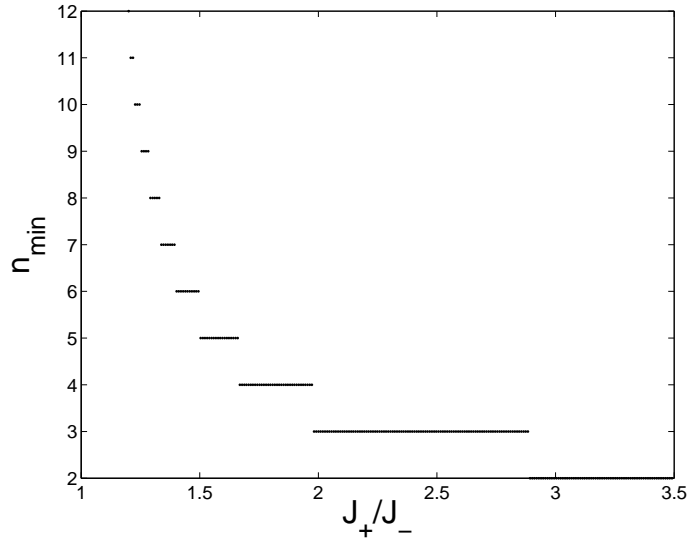}
\caption{\sf The minimum allowed degree $n_{\rm min}$ of a quasibreather
as a function of the inhomogeneity parameter $J_+/J_-$.}
\label{fig:gmka}
\end{figure}

Recall that the trajectories depicted in figure \ref{fig:diw} are
the projections of geodesics to the $(\sigma,p_\sigma)$ plane. A closed
trajectory in this picture does not necessarily correspond to a periodic
geodesic: both the width $\sigma(t)$ and the phase $e^{i\psi(t)}$ are
 periodic, but with (in general)
incommensurate periods. These geodesics are therefore only
quasiperiodic, so we call the oscillating $n$-bubbles
associated with them quasibreathers.
The time dependence of the width $\sigma$ and internal phase
${\psi}$ of a typical quasibreather is depicted in figure 
\ref{fig:dnt}.
Note that the shape of these solutions
really does ``breathe''. This is in contrast to the internally spinning
magnetic solitons (sometimes called breathers) which occur in
{\em ferromagnetic}
spin lattices (with on-site anisotropy),
where $n_3$ has a fixed hump-shaped profile and the
spins merely
rotate at constant speed about the $n_3$ axis \cite{kosivakov}.
For $n=2$, quasibreathers exist only for $J_+/J_->2.89$ (approximately),
but for any value of $J_+/J_->1$, there is some $n_{\rm min}\geq 2$ such
that the geodesic flow supports quasibreathers of any degree $n\geq n_{\rm
min}$. The dependence of $n_{\rm min}$ on $J_+/J_-$ is shown in figure
\ref{fig:gmka}.

\section{Concluding remarks}

We have seen that, in the near continuum regime, a two-dimensional
isotropic Heisenberg antiferromagnet with position-dependent
exchange integral
is described by the relativistic $O(3)$ sigma model on a spatially
inhomogeneous spacetime. By conformal invariance, such a model supports
static Belavin-Polyakov lumps, which we call bubbles. The trajectories
of these bubbles are predicted to behave like light rays propagating
in an inhomogeneous medium, with $J^{-1}$, the inverse exchange integral,
identified with the refractive 
index. In particular, bubbles incident on a straight domain wall are
refracted in accordance with Snell's law, and total internal reflexion can
occur. This refraction phenomenon was noted in \cite{spebub}.

In this paper we have reviewed the analysis of bubble
dynamics presented in \cite{spebub}, and extended it beyond
the scattering problems previously considered to include the
rotationally equivariant dynamics of higher degree bubbles. Here we found
that disk $J$-inhomogeneities with sufficiently large $J_+/J_-$
support quasibreathers: degree $n\geq 2$ bubbles centred in the disk
which spin internally while their shape oscillates with a
generically incommensurate period.

Our analysis has been conducted entirely  within the framework of 
the geodesic approximation of Manton.
The validity of this approximation for the $O(3)$ sigma model
remains an open question. In the general context of relativistic field
theories of Bogomol'nyi type, it has proved to be
very succesful at describing soliton scattering 
processes at low to intermediate energies, where the solitons move 
essentially as free particles except during a relatively brief interaction
phase \cite{mansut}.
The basic domain wall refraction effect seems likely to be a robust 
feature of the degree 1 bubble dynamics, therefore.

The status of the quasibreather solutions is less straightforward.
It is unlikely that the sigma model will support genuine
quasiperiodic solutions for any bounded choice of $J(|z|)$, because one
expects an oscillating, spinning bubble to radiate energy to infinity
in the form of small amplitude travelling waves. 
So the quasiperiodic geodesics in $\mset_n$ 
found here cannot be globally close to the 
true solutions of the model with the same initial data. 
One should
think of the geodesic approximation as predicting that long-lived
oscillating, spinning bubbles exist in the model which are approximately
quasiperiodic, but decay over a long time scale. 

The only
geodesics for which rigorous error estimates have been found are the
radial half rays $\rho(t)=-vt$ in $\mset_n$ for $n\geq 3$
(in the homogeneous case, $J$ constant, though this is unlikely to be
crucial). These geodesics
describe an equivariant $n$-bubble collapsing in finite time.
Rodnianski and Sterbenz \cite{rodste} have shown that the actual
solutions do indeed collapse in finite time, but that the collapse
rate predicted by the geodesic approximation receives a logarithmic
correction which becomes significant when the bubble is extremely
close to collapse (i.e.\ extremely narrow). In the present context, the
significance of this deviation is debatable: when the bubble is
very narrow, the continuum approximation which led to the sigma model in the
first place has presumably already broken down. Indeed, the most important
test of the dynamical phenomena predicted here and in
\cite{spebub} is not provided by
rigorous analysis of the PDE (\ref{oli}), but by direct numerical simulation
of the spin lattice (\ref{dag}) itself. Such simulation is computationally
far more intensive than the methods we have used here, and lies beyond the 
scope of this paper.

\section*{Acknowledgements}

This paper is based largely on the talk
``Magnetic bubble refraction in inhomogeneous antiferromagnets''
given by the author at the workshop
Nonlinear Physics Theory and Experiment IV, Gallipoli, Italy 2006.\,  
 His attendance at this workshop 
was funded by the EPSRC.

\end{document}